\documentclass{article}
\usepackage{graphicx}
\begin{document}



%
%

\title{Generalized Exponential Function and some of its Applications to Complex Systems}

\author{ Alexandre Souto Martinez\thanks{asmartinez@ffclrp.usp.br},\\  
             Universidade de S\~ao Paulo \\ 
             Faculdade de Filosofia, Ci\^encias e Letras de Ribeir\~ao Preto \\
             Avenida Bandeirantes, 3900 \\ 
             14040-901, Ribeir\~ao Preto, S\~ao Paulo, Brazil.\\
             and \\
             National Institute of Science and Technology for Complex Systems
\\ \\
 Rodrigo Silva Gonz\'alez\thanks{caminhos\_rsg@yahoo.com.br} and 
             C\'esar Augusto Sangaletti Ter\c{c}ariol\thanks{cesartercariol@gmail.com} \\ 
             Universidade de S\~ao Paulo \\ 
             Faculdade de Filosofia, Ci\^encias e Letras de Ribeir\~ao Preto \\
             Avenida Bandeirantes, 3900 \\ 
             14040-901, Ribeir\~ao Preto, S\~ao Paulo, Brazil.
             }

\maketitle


\begin{abstract}
From the integration of non-symmetrical hyperboles, a one-parameter generalization of the logarithmic function is obtained.
Inverting this function, one obtains the generalized exponential function.
We show that functions characterizing complex systems can be conveniently written in terms of this generalization of the exponential function.
The gamma function is then generalized and we generalize the factorial operation. 
Also a very reliable rank distribution can be conveniently described by the generalized exponential function. 
Finally, we turn the attention to the generalization of one- and two-tail stretched exponential functions. 
One obtains, as particular cases, the generalized error function, the Zipf-Mandelbrot probability density function (pdf), the generalized gaussian and Laplace pdf.
One can also obtain analytically their cumulative functions and moments.
\\
{\bf Keywords:} generalization of exponential function; generalization of the factorial operation; generalization of the stretched exponential function.
\end{abstract}


\section{Introduction}

The convenience of generalizing the logarithmic function has attracted the attention of researchers since long ago~\cite{rogers_1894} and particularly in the last years~\cite{bienharn_1989,macfarlane_1989,floreanini_1995,mcanally_1995a,mcanally_1995b,atakishiyev_1996a}. 
In Physics, several one-parameter generalizations of the logarithmic function have been proposed in different contexts such as: 
non-extensive statistical mechanics~\cite{tsallis_1988,tsallis_qm,nivanen_2003,borges_2004,kalogeropoulos_2005}, relativistic statistical mechanics~\cite{kaniadakis_2001,PhysRevE.66.056125} and 
quantum group theory~\cite{abe_1997}. 
Also, more sophisticated  such as two-parameter~\cite{kaniadakis:046128} and three-parameter~\cite{kaniadakis:036108} generalizations have been proposed, each one including previous situations as particular cases. 
Examples of the convenience of these generalizations have been seen in different fields, as for instance: 
psychophysics~\cite{takahashi_2_2008}, 
neuroeconomics~\cite{takahashi_2_2007,cajueiro_2006},
econophysics~\cite{takahashi_3_2007,anteneodo:1:2002}, 
complex networks~\cite{holanda:2004,albuquerque:2000},
population dynamics~\cite{martinez:2008a,martinez:2008b} etc. 

Here, our main objective is to show that the generalized the stretched exponential function, written as probability density function (pdf),  is convenient to describe complex systems. 
In Sec.~\ref{sec:review}, we show that from the integration of non-symmetrical hyperboles, one obtains a one-parameter generalization of the logarithmic function, which we call $\tilde{q}$-logarithm.
This generalization coincides to the one obtained in the context of non-extensive thermostatistics~\cite{tsallis_qm}. 
Inverting the $\tilde{q}$-logarithm, one obtains the generalized exponential ($\tilde{q}$-exponential) function. 
Some properties of these generalized functions are presented.
In Sec.~\ref{app:generalized_gamma}, the gamma function is generalized and this allows us to generalize the factorial operation, which does not match the previous definition~\cite{suyari_2006,arruda_2008}, but as we show, it is self consistent. 
In Sec.~\ref{sec:rank_dist}, the very reliable rank distribution obtained by Naumis and Cocho~\cite{naumis-2007} is conviently described by the $\tilde{q}$-exponential function stressing finite size effects. 
In Sec.~\ref{sec:special}, we first show that the Zipf-Mandelbrot function, which is a fingerprint of complex systems, can be conveniently written in terms of the $\tilde{q}$-exponential. 
Raising the $\tilde{q}$-exponential argument to a given power, one obtains a function that generalizes the stretched exponential function. 
Its generating differential equation is then presented. 
In Sec.~\ref{sec:prob_dist}, we consider the pdf's for continuous variables.  
First we consider the one-tail stretched exponential generalization and obtain analytically its cumulative function and moments. 
One obtains the generalized error function as a particular case and the Zipf-Mandelbrot pdf as another. 
Next, we consider the two-tail generalized stretched exponential pdf and obtain analytically its cumulative function and moments. 
One has the generalized gaussian and the generalized laplace pdf as particular cases. 
The characteristic function is analitically calculated.  
Our final remarks are drawn in Sec.~\ref{sec:conclusion}.

\section{The $\tilde{q}$-generalized functions}
\label{sec:review}

From the integration of non-symmetrical hyperboles, we obtain a one-parameter generalization of the logarithmic function, which coincides to the one obtained in the context of non-extensive thermostatistics~\cite{tsallis_qm}. 
Inverting this function, one obtains the generalized exponential function.
Some properties of these generalized functions are presented.

\subsection{$\tilde{q}$-generalized logarithm function}

In the one-parameter generalization we address here, the $\tilde{q}$-logarithm function $\ln_{\tilde{q}}(x)$ is defined as the value of the area  underneath $1/t^{1-\tilde{q}}$, in the interval $t \in [1,x]$~\cite{arruda_2008}:
\begin{eqnarray}
\ln_{\tilde{q}}(x) & = & \int_1^x \frac{\mbox{d}t}{t^{1-\tilde{q}}} = \lim_{\tilde{q}' \rightarrow \tilde{q}}\frac{x^{\tilde{q}'} - 1}{\tilde{q}'} \; . 
\label{eq:gen_log}
\end{eqnarray}
This is exactly the same function obtained from the non-extensive statistical mechanics context~\cite{tsallis_1988,tsallis_qm}, which uses $q = 1 - \tilde{q}$. 

The usual natural logarithm ($\ln x$) is retrieved for $\tilde{q} = 0$ and a linear function for $\tilde{q} = 1$. 
Scaling and deformation of the variable $x$ are given by: $\ln_{\tilde{q}}(\alpha x^{\beta}) = \beta \ln_{\beta \tilde{q}}(\alpha^{1/\beta} x)$, so that for  $\beta = -1$, one has 
$\ln_{\tilde{q}}(\alpha/x) = - \ln_{- \tilde{q}}(x / \alpha)$ and for the particular case $\alpha = 1$, $ \ln_{\tilde{q}}(x^{-1}) = - \ln_{-\tilde{q}}(x)$.

\subsection{$\tilde{q}$-generalized exponential function}

The $\tilde{q}$-exponential function $e_{\tilde{q}}(x)$ is defined as the $t$-value, in such a way that the area underneath $f_{\tilde{q}}(t) = 1/t^{1-\tilde{q}}$, in the interval $t \in [1,e_{\tilde{q}}(x)]$, is $x$.
This is the inverse of the $\tilde{q}$-logarithm function $e_{\tilde{q}}[\ln_{\tilde{q}}(x)] = x = \ln_{\tilde{q}}[e_{\tilde{q}}(x)]$ and it is given by: 
\begin{eqnarray}
e_{\tilde{q}}(x) & = & \left\{ 
                       \begin{array}{ll}
                       0 & \mbox{for} \; \tilde{q} x < -1 \\  
                       \lim_{\tilde{q}' \rightarrow \tilde{q}}(1 + \tilde{q}'x)^{1/\tilde{q}'} & 
                       \mbox{for} \; \tilde{q} x \ge -1\end{array}
                       \right. \; .
\label{eq:q_exp}
\end{eqnarray}
Notice that $(1 + \tilde{q}x)^{1/\tilde{q}}$ is real only if $\tilde{q} x \ge -1$.  

This is a non-negative function [$e_{\tilde{q}}(x) \ge 0$, for all $\tilde{q}$ and $x$] and $e_{\tilde{q}}(0) = 1$, independently of the $\tilde{q}$ value.
For $\tilde{q} = 0$, one retrieves the usual exponential function $e^x$ and for $\tilde{q} = 1$, a linear function.
Notice that taking the surface underneath $f1/t^{1-\tilde{q}}$ to be unitary, one  generalizes the Euler's number: $e_{\tilde{q}} = e_{\tilde{q}}(1) = (1 + \tilde{q})^{1/\tilde{q}}$. 
An interesting property is that $[e_{\tilde{q}}(x)]^a = e_{\tilde{q}/a}(a x) $, meaning that  the $\tilde{q}$-exponential argument scaling corresponds to a power of a different $\tilde{q}$-exponential function.   
For $a = -1$, one has that $e_{-\tilde{q}}(-x) = 1/e_{\tilde{q}}(x)$. 
 
The derivative of the  $\tilde{q}$-exponential function with respect to $x$ is:
$\mbox{d} e_{\tilde{q}}(k x)/\mbox{d} x = k \; [e_{\tilde{q}}(k x)]^{1-\tilde{q}} = k \; e_{\tilde{q}/(1-\tilde{q})}[(1-\tilde{q}) k x]$, so that it is the solution of the following non-linear first order differential equation: $\mbox{d} y(x)/\mbox{d}x = k y^{1-\tilde{q}}(x)$, which is a particular case of Bernoulli's differential equation: 
$[\mbox{d}/\mbox{d}x  + p(x) ] y(x) = q(x) y^{1-\tilde{q}}(x)$, with $p(x) = 0$ and $q(x) = k$. 
Notice that $k$ has the dimension of $y^{\tilde{q}}$ over the dimension of $x$. 
This means that it sets up a scale (inversion dimension of $x$) to the problem only if $\tilde{q} = 0$.
One important application of the above differential equation concerns reaction kinetics~\cite{niven_2006}.

The successive derivatives of the $\tilde{q}$-exponential function with respect to $x$ are written as:
\begin{eqnarray}
\frac{\mbox{d}^n e_{\tilde{q}}(x)}{\mbox{d} x^n} & = & \left[ \prod_{k=0}^{n-1}(1-k\tilde{q}) \right] \; \left[ e_{\tilde{q}}(x) \right]^{1- n \tilde{q}} 
                                               = \frac{(-\tilde{q})^{n} \Gamma(n - 1/\tilde{q} )}{\Gamma(-1/\tilde{q})} \; \left[ e_{\tilde{q}}(x) \right]^{1- n \tilde{q}} \; ,
\label{eq:sucessive_derivatives_eq}
\end{eqnarray}
where to get a closed form, one has used :
$\prod_{k = 0}^{n-1}(1-k\tilde{q}) = (-\tilde{q})^{n} \Gamma(n - 1/\tilde{q} )/\Gamma(-1/\tilde{q})$, which can be verified by induction. 
As far as we are aware, this is a new closed analytical form. 

The $\tilde{q}$-exponential function can be developed in Taylor's series around $x = 0$:
\begin{eqnarray}
e_{\tilde{q}}(x) & = & \sum_{n=0}^{\infty} \left. \frac{\mbox{d}^n e_{\tilde{q}}(x)}{\mbox{d} x^n} \right|_{x=0} \; \frac{x^n}{n!} 
                  = \sum_{n=0}^{\infty} \frac{(-\tilde{q})^{n} \Gamma(n - 1/\tilde{q} )x^n}{\Gamma(-1/\tilde{q})n!} \; , 
\label{eq:taylor}
\end{eqnarray}
where we have used that $e_{\tilde{q}}(0) = 1$ in Eq.~\ref{eq:sucessive_derivatives_eq}. 

\section{Generalized Factorial}
\label{app:generalized_gamma}

The $\tilde{q}$-gamma function is defined as:
\begin{equation}
\Gamma_{\tilde{q}}(a) =  \int_0^{1/\tilde{q}} dt \; t^{a-1}e_{\tilde{q}}(-t) \; ,
\end{equation}
with $a > 0$. 
For $\tilde{q} \to 0$, one retrieves the standard gamma function~\cite{stegun}, since $e_0(-t) = e^{-t}$ and $1/\tilde{q} \to \infty$. 
For $\tilde{q} > 0$, one has: $\Gamma_{\tilde{q}}(a \to 0) =  1/a$, 
 $\Gamma_{\tilde{q}}(1) = \int_0^{1/\tilde{q}} dt \; e_{\tilde{q}}(-t) = 1/(1 + \tilde{q})$ etc. 
 
We point out that $\tilde{q}$-generalized gamma function is closely related to the beta function 
$\Gamma_{\tilde{q}}(a) = \mbox{B}(a,1/\tilde{q} + 1)/\tilde{q}^a$.
From the properties of the beta function, one obtains the following recursion relations for the $\tilde{q}$-gamma function:
\begin{equation}
\Gamma_{\tilde{q}}(a+1) = \frac{a \; \Gamma_{\tilde{q}}(a)}{1 + \tilde{q}(a+1)}  \; 
\label{eq:recurrence_q_gamma}
\end{equation}
which generalizes the factorial operation: 
\begin{equation}
(a) _{\tilde{q}}! = \frac{a \; (a - 1)_{\tilde{q}}!}{1 + \tilde{q}(a+1)} \; ,
\end{equation}
so that $(a \to 0) _{\tilde{q}}! = 1/a$. 
This definition does not match the one previously proposed for integer values~\cite{suyari_2006,arruda_2008}: $n !_{\tilde{q}}  = e_{\tilde{q}} \left[ \sum_{k=1}^n \ln_{\tilde{q}}(k) \right]$. 

As stressed by Borges~\cite{borges_1998}, other possible generalizations of the exponential function are defined as one writes the generalization of the factorial in their Taylor expansion. 
Writing the Taylor expansion of the $\tilde{q}$-exponential function around $x=0$ (Eq.~\ref{eq:taylor}) as:
\begin{eqnarray}
\nonumber
e_{\tilde{q}}(x) & = & \sum_{n=0}^{\infty}  \frac{x^n}{(n)_{-\tilde{q}}!} \; , 
\end{eqnarray}
we define the $\tilde{q}$-factorial as: 
\begin{equation}
n _{-\tilde{q}}!  = \frac{-\Gamma(-1/\tilde{q}) \; n! }{(-\tilde{q})^{n} \Gamma(n - 1/\tilde{q} )} = \frac{n(n-1) _{-\tilde{q}}!}{1 - \tilde{q}(n-1)} \; .
\label{eq:q_fatorial_taylor}
\end{equation}

This definition of the generalization of the factorial is compatible with the generalization of the factorial obtained from the generalization of the gamma function if one write:
\begin{equation}
n _{\tilde{q}}! = \frac{n(n-1) _{\tilde{q}}!}{1 + \tilde{q}[n + \mbox{sgn}(\tilde{q})]} \; . 
\label{eq:q_fatorial_taylor_geral}
\end{equation}
where $\mbox{sgn}(x) = x/|x|$.

For real and complex values of $a$, we propose the following definition for the generalized factorial function:
\begin{equation}
a _{\tilde{q}}! = \frac{a \; (a - 1) _{\tilde{q}}!}{1 + \tilde{q}[a+ \mbox{sgn}(\tilde{q})]} \; .
\end{equation}

\section{Beta-like Distribution}
\label{sec:rank_dist}

Let us turn our attention to discrete random variables, the rank distribution in particular.  
We show that the rank distribution obtained by Naumis and Cocho~\cite{naumis-2007} is conviently described by the $\tilde{q}$-exponential function and finite size effects are stressed. 
This rank distribution is very reliable since it has an underlining microscopic model and has been validated by a wide range of experimental data~\cite{naumis-2007}. 

To simultaneously fit the beginning, body and tail of experimental rank distributions of complex systems, Naumis and Cocho~\cite{naumis-2007} consider $N$ independent subsystems with a large number of internal states.
The rank $r$ of a system property dependent on the internal states of the subsystems decays as a two-free-parameter beta-like function: $f(r)  =  K [1 - r/(R + 1) ]^{\beta}r^{\alpha}$, where $K    =  1/\sum_{r = 1}^N [1 -r/(R + 1)]^{\beta} r^{-\alpha}$,  with $R \le N$ is the maximal value of $r$, $K$ is the normalization factor and the two free parameters are $\alpha$ and $\beta$. 
As noticed by the autors, if $R \gg 1$, then $K \approx 1/\mbox{B}(1-\alpha,1 + \beta)$.

Finite size effects are described by factor $[1 - r/(R + 1)]^{\beta} \approx 1 - \beta r/R \approx e^{- r/r_0}$, for $R \gg r$ with $r_0 = R/\beta$. 
In this way, we see that the rank distribution of $f(r)$ is in fact a generalization of the standard technique of multiplying the power-law ($r^{-\alpha}$) by an exponential cutoff: $f(r) \propto r^{-\alpha} e^{- r/r_0}$.  

If one writes: $[1 - r/(R + 1)]^{\beta} = [1 - (1/\beta) \beta r/(R + 1)]^{\beta} = e_{1/\beta}[- \beta r/(R+1)]$ then 
\begin{equation}
f(r)  =  \frac{K}{r^{\alpha}} \; e_{1/\beta}\left[ \frac{-\beta r}{R+1} \right] = \frac{K}{r^{\alpha}} \; \left[e_{1}\left( \frac{- r}{R+1} \right)\right]^{\beta} \; , 
\end{equation}
meaning that the $\tilde{q}$-exponential function can properly take into account finite size effects.
The exponential cutoff is retrieved where $R$ and $b$ grow, but $R$ must grow faster than $\beta$, to have $R \rightarrow \infty$ and $\beta \rightarrow \infty$, but $\beta/R \rightarrow 0$. 

\section{Special Functions and Processes}
\label{sec:special}

In what follows, we first show that the Zipf-Mandelbrot function, which is a fingerprint of complex systems, can be written in terms of the $\tilde{q}$-exponential. 
Next, raising the $\tilde{q}$-exponential argument to a given power, one obtains a function that generalizes the stretched exponential function. 
Finally, we obtain the process (differential equation) of which the generalized stretched exponential is the solution.

\subsection{Zipf-Mandelbrot Function}

The envelope of a rank distribution  of complex systems can sometimes be very well described by the Zipf-Mandelbrot function~\cite{picoli_2005}, which may be written in terms of $\tilde{q}$-exponential function: 
\begin{eqnarray}
P_{\tilde{q}, \alpha, A}(x) & = & \frac{d}{(c + x)^{\gamma}} = \frac{A}{e_{\tilde{q}}(x/\alpha)} 
             = A \; e_{-\tilde{q}}\left(\frac{- x}{\alpha} \right) \; ,
\label{eq:zipf_mandelbrot_function}
\end{eqnarray}
with $\tilde{q} = 1/\gamma$, $\alpha = c/\gamma$ and $A = d/c^{\gamma}$. 

We remark that Eq.~\ref{eq:zipf_mandelbrot_function} has another interesting application in time-dependent luminescence spectroscopy.
In this case the relaxation processes are known as Becquerel decay function~\cite{Berberan-Santos_2005:b}. 

\subsection{Generalized Stretched Exponential Function}

In some instances, to model some characteristic $x$ of a complex system it is necessary a deformation in the argument of Eq.~\ref{eq:zipf_mandelbrot_function}, which leads to the generalized stretched exponential function: 
\begin{eqnarray}
P_{\tilde{q}, \alpha, \beta, A}(x) & = & P_{\tilde{q}, \alpha, A}(x^{1/\beta}) = \frac{d}{(d + x^{1/\beta})^{\gamma}} 
 = A \; e_{-\tilde{q}}\left( \frac{- x^{1/\beta}}{\alpha} \right) \; .
\label{eq:gen_stretched_exp_function}
\end{eqnarray}

The usual stretched exponential function, also known as the Kohlrausch function~\cite{cardona_2007,Berberan-Santos_2008:a}, is obtained from Eq.~\ref{eq:gen_stretched_exp_function} in the limit $\tilde{q} \to 0$. 
Although the (usual) stretched exponential function has been used to describe relaxation processes in time-dependent luminescence spectroscopy~\cite{Berberan-Santos_2008:a}, a generalization of the form $ e^{\beta/a} \; e^{-e_{1/\beta}(x/\alpha)/a}$ seems to be more convenient to fit experimetal data~\cite{Berberan-Santos_2008:b}.
In this case we stress that the $\tilde{q}$-exponential figures as the argument of the usual exponential function.


The stretched exponential function can also be obtained as the solution of the following differential equation: $dy(x)/dx = - x^{1/\beta - 1} y(x)/(\alpha\beta)$, which can be written in terms of the logarithmic function as: 
$d\ln y(x)/dx = - x^{1/\beta - 1}/(\alpha\beta)$. 
If one replaces the logarithmic function in the differential equation obtained above by the $\tilde{q}$-logarithm, one obtains:
\begin{equation}
\frac{d \, \ln_{-\tilde{q}} y(x)}{dx} = - \; \frac{x^{1/\beta - 1}}{\alpha\beta} \; , 
\end{equation}
or equivalently
\begin{equation}
\frac{dy(x)}{dx} = - \; \frac{x^{1/\beta - 1}}{\alpha\beta} y^{1+\tilde{q}}(x) \; , 
\end{equation}
which is a particular case of Bernoulli's equation: $d y(x)/dx + p(x) y(x) = q(x)y^{1+\tilde{q}}(x)$, with $p(x)=0$ and $q(x)=x^{1/\beta-1}/(\alpha\beta)$ and which solution is precisely the generalized stretched exponential function of Eq.~\ref{eq:gen_stretched_exp_function}.

\section{Probability Functions}
\label{sec:prob_dist}

Considering the factor $A > 0$ and using the non-negativeness of Eq.~\ref{eq:gen_stretched_exp_function}, we write the generalization of the stretched exponential pdf and study its properties. 
We consider one- and two-tail distributions and obtain some known pdf's (generalized gaussian) and new ones (generalized error function and generalized Laplace pdf) as particular cases.  

\subsection{One-Tail PDF}
\label{sec:one_tail}

If the considered independent variable $x$ is constrained to non-negative (or eventually non-positive) values, then one uses the one-tail pdf. 
We consider the generalization of the stretched exponential pdf and analytically obtain its cumulative function and moments. 
From this pdf, one obtains the generalized error function as a particular case and the Zipf-Mandelbrot pdf as another. 
 
The normalization factor $A$ of Eq.~\ref{eq:gen_stretched_exp_function} is: 
\begin{eqnarray}
\frac{1}{A} = \int_{0}^{\infty} dt \; e_{- \tilde{q}} \left(\frac{- t^{1/\beta}}{\alpha} \right) = 
\beta \left( \frac{\alpha}{\tilde{q}} \right)^{\beta} \mbox{B} \left(\beta, \frac{1}{\tilde{q}} - \beta \right) \; ,
\label{Eq:Normaliz}
\end{eqnarray}
where $\mbox{B}(a,b) = \Gamma(a)\Gamma(b)/\Gamma(a+b)$ is the beta function~\cite{stegun} and $\Gamma(x) = \int_0^{\infty} dt \; t^{x-1} e^{-t}$, $x > 0$ is the gamma function~\cite{stegun}.
The integral of Eq.~\ref{Eq:Normaliz} does not diverge only if $0<\tilde{q}<1$, $0<\beta<1/\tilde{q}$, and $\alpha>0$ and one has the \emph{generalized stretched exponential pdf}:
\begin{eqnarray}
P_{\tilde{q}, \alpha, \beta}^{(1)}(x)& = & 
\frac{(\tilde{q}/\alpha)^{\beta}}{\beta \mbox{B}(\beta,1/\tilde{q} - \beta)}  \;  
     e_{-\tilde{q}} \left[ - \frac{x^{1/\beta}}{\alpha} \right] \; . 
\label{eq:generalized_strechted_exponential_0}
\end{eqnarray}

\begin{figure}[htb!]
 \begin{center}
 \includegraphics[width=.65\columnwidth]{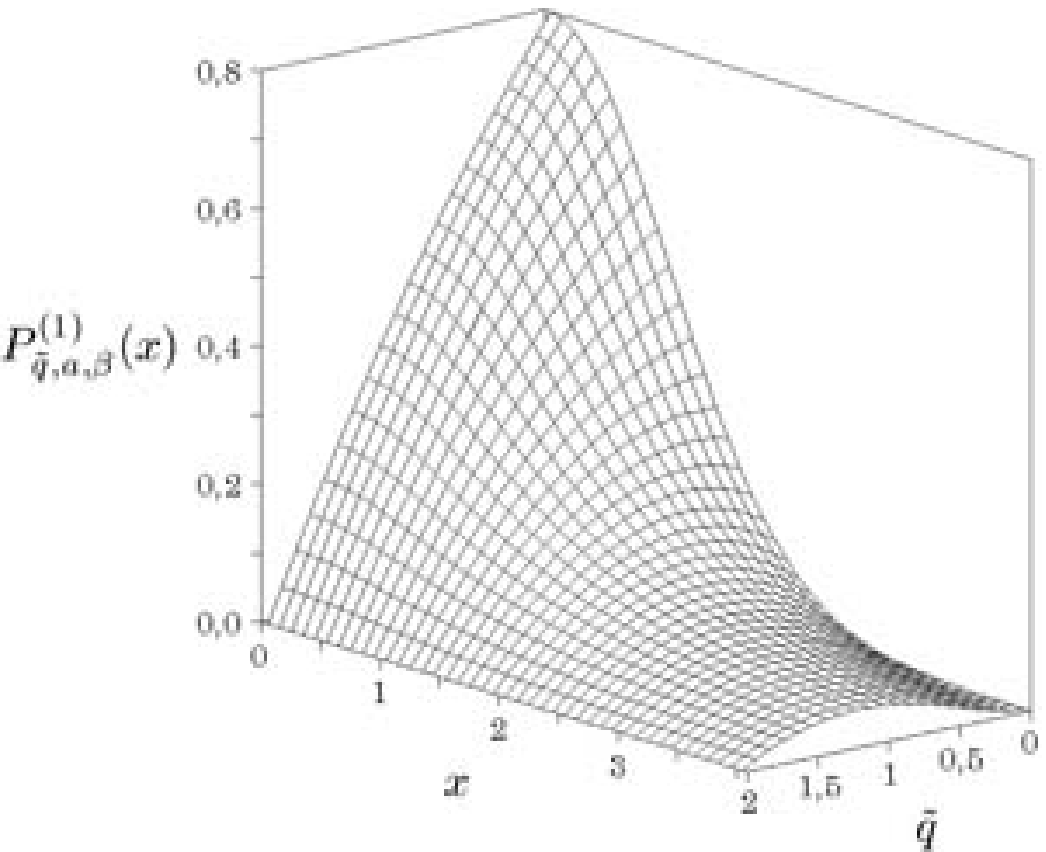}
 {\bf (a)}
 \includegraphics[width=.65\columnwidth]{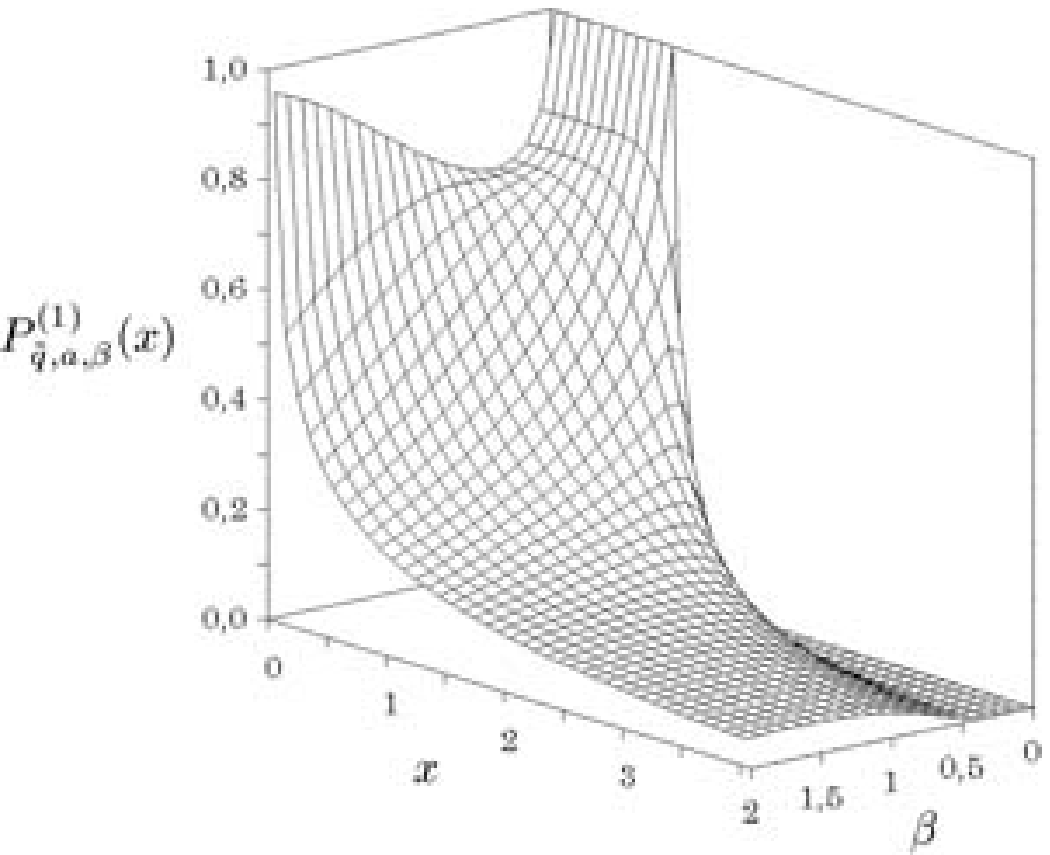}
 {\bf (b)}
 \includegraphics[width=.65\columnwidth]{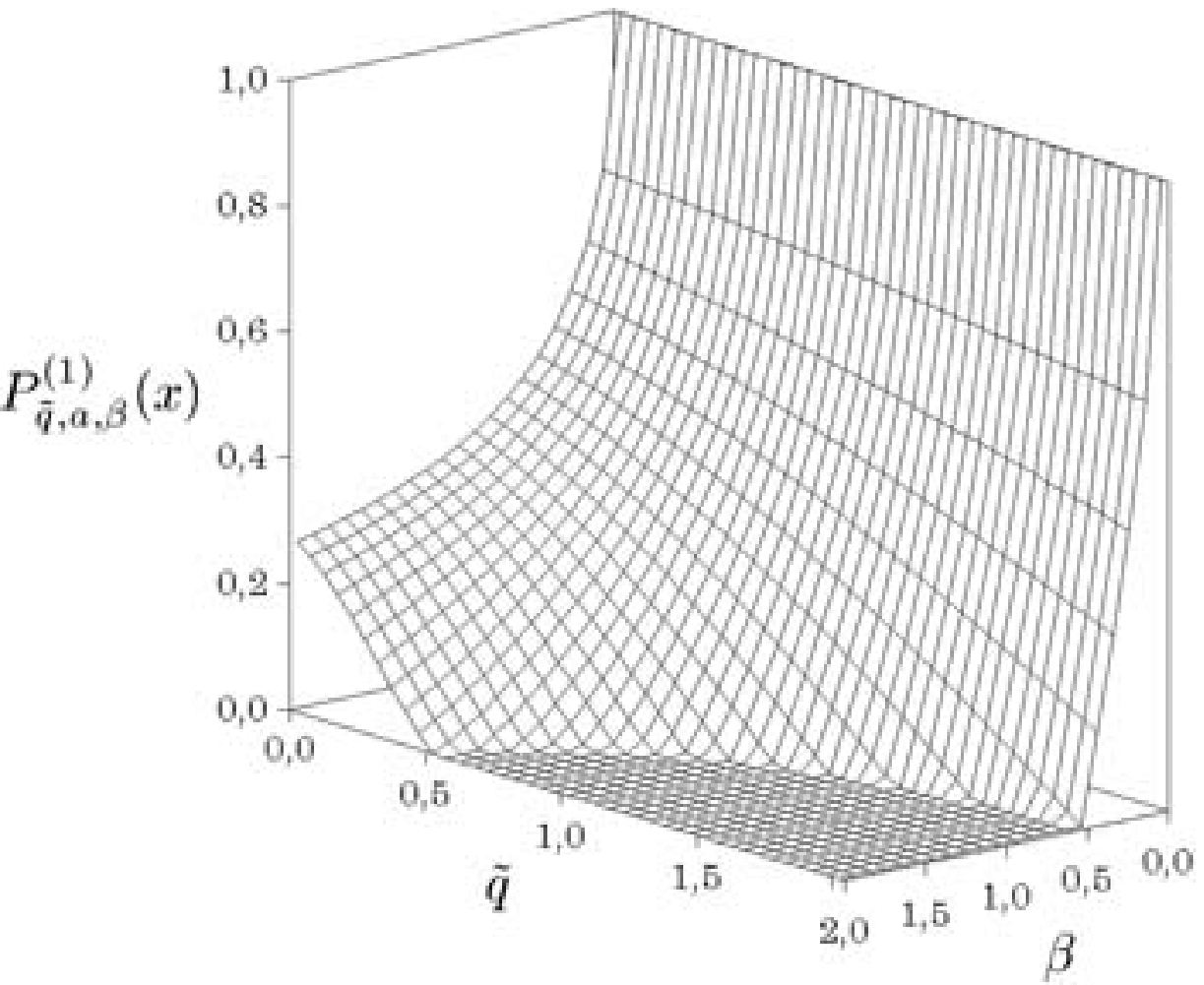}
 {\bf (c)} 
 \end{center}
 \caption{Behavior of Eq.~\ref{eq:generalized_strechted_exponential} as a function of: 
          {\bf (a)}  $x$ and $\tilde{q}$, with $\beta = 1/2$ and $a = 1$,  
          {\bf (b)}  $x$ and $\beta$, with $\tilde{q} = 1/5$ and $a = 1$,
          {\bf (c)}  $\tilde{q}$ and $\beta$, with $x = 1$ and $a = 1$.          
         } 
 \label{fig:zipft_onetail:1}
 \end{figure}

\begin{figure}[htb!]
 \begin{center}
 \includegraphics[width=.95\columnwidth]{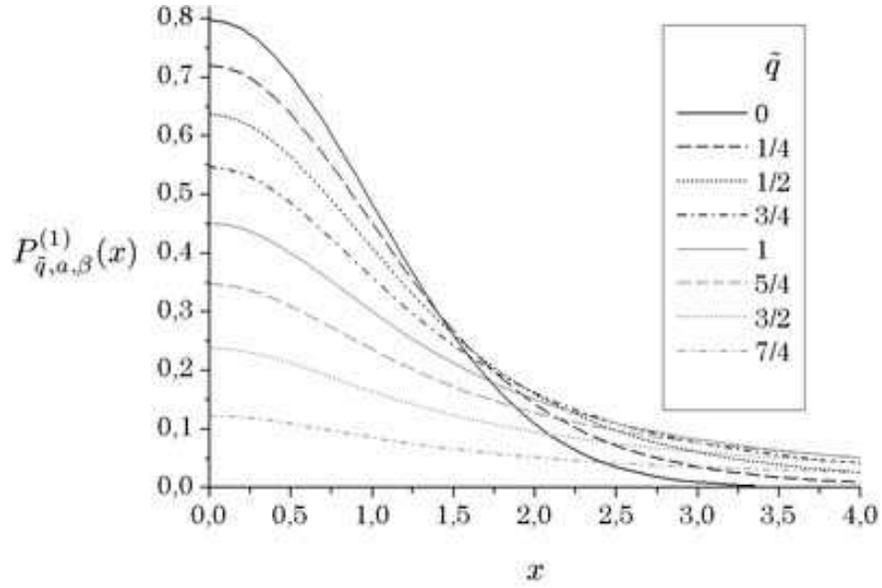}
 {\bf (a)}
 \includegraphics[width=.95\columnwidth]{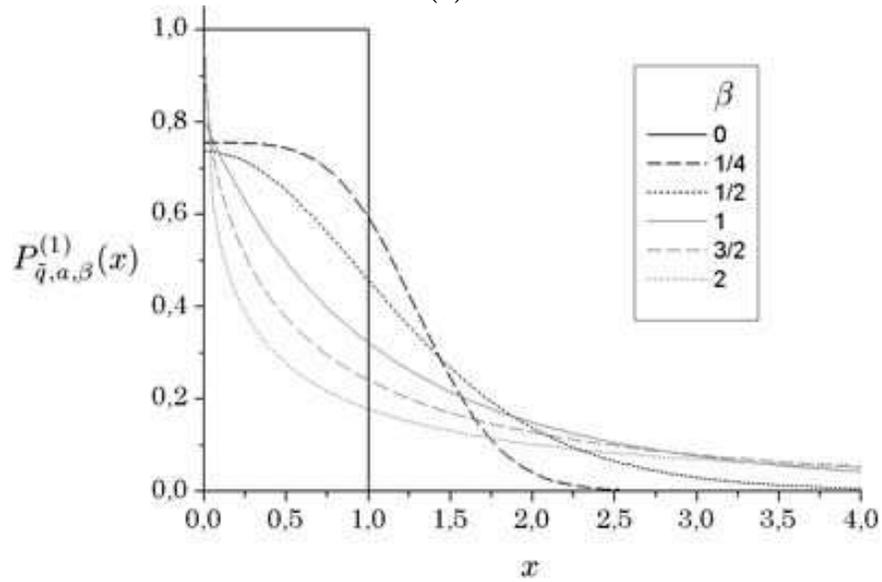}
 {\bf (b)}
 \end{center}
 \caption{Behavior of Eq.~\ref{eq:generalized_strechted_exponential} as a function of: 
          {\bf (a)}  $x$  with $\beta = 1/2$ and $a = 1$ and for several $\tilde{q}$ values (Projection of Fig.~\ref{fig:zipft_onetail:1}a), 
          {\bf (b)}  $x$  with $\tilde{q} = 1/5$ and $a = 1$ and for several $\beta$ values (Projection of Fig.~\ref{fig:zipft_onetail:1}b). 
         } 
 \label{fig:zipft_onetail:2}
 \end{figure}

For our purposes, it is more convenient to write $a = (\alpha\beta)^\beta > 0$:
\begin{eqnarray}
P_{\tilde{q},a,\beta}^{(1)}(x) & = & 
\frac{ (\tilde{q} \beta )^{\beta}} 
     {a \beta \mbox{B}(\beta,1/\tilde{q} - \beta)}  \;  
     e_{-\tilde{q}} \left[ - \beta \; \left( \frac{x}{a} \right)^{1/\beta} \right] \; , 
\label{eq:generalized_strechted_exponential}
\end{eqnarray}
which is depicted in Figs.~\ref{fig:zipft_onetail:1} and~\ref{fig:zipft_onetail:2}. 

As $\tilde{q} \to 0$, $\mbox{B}(\beta, 1/\tilde{q} - \beta) \approx \tilde{q}^{\beta} \Gamma(\beta)$, since $1/\tilde{q} \gg \beta$, one retrieves the stretched exponential function $P_{0,a,\beta}^{(1)}(x) = \beta^{\beta} e^{-\beta(x/a)^{1/\beta}}/[a \beta \Gamma(\beta)]$ or $P_{0, \alpha, \beta}^{(1)}(x) = e^{-x^{1/\beta}/\alpha}/[\alpha^\beta \beta \Gamma(\beta)]$.

The cumulative function of Eq.~\ref{eq:generalized_strechted_exponential} is: 
\begin{eqnarray}
F_{\tilde{q},a,\beta}^{(1)}(x) & = & \int_{0}^{x} dt \; P_{\tilde{q},a,\beta}^{(1)}(t) \nonumber 
= \frac{x}{A} \; _2\mbox{F}_1 \left( \beta, \frac{1}{q}, \beta+1; - \frac{\tilde{q} x^{1/\beta}}{\alpha} \right)
\label{eq:cumulative_generalized_strechted_exponential}
\end{eqnarray}
where 
\begin{equation}
_2\mbox{F}_1(a, b; c; x) = \sum_{n=0}^\infty \frac{(a)_n (b)_n}{(c)_n} \frac{x^n}{n!}
\label{Eq:HypGeomFun}
\end{equation}
is the hypergeometric function~\cite{stegun} and
\begin{eqnarray*}
(a)_n & = & a(a+1)(a+2)\cdots(a+n-1) 
= \frac{(a+n-1)!}{(a-1)!} = \frac{\Gamma(a+n)}{\Gamma(a)}
\end{eqnarray*}
is the Pochhammer symbol.

The moments of $x$ are:
\begin{eqnarray}
\langle x^n \rangle_{\tilde{q},a,\beta} = \left[ \frac{a}{(\tilde{q} \beta)^{\beta}} \right]^{n} \frac{\mbox{B}[(n+1) \beta, 1/\tilde{q} - (n+1)\beta]}{\mbox{B}(\beta,1/\tilde{q} - \beta)} \; , 
\end{eqnarray}
where one sees they are finite only if $\tilde{q} < 1/[(n+1) \beta]$. 

%

If $\tilde{q} < 1/(3 \beta)$, the mean value and variance are finite and, respectively, given by:
\begin{eqnarray}
\langle x \rangle_{\tilde{q},a,\beta} & = & \left(\frac{\alpha}{\tilde{q}}\right)^{\beta} \;
\frac{\mbox{B}(2 \beta,1/\tilde{q} - 2 \beta)}
     {\mbox{B}(  \beta,1/\tilde{q} -   \beta)}    \\
\frac{\sigma_{\tilde{q},a,\beta}^2}
     {\langle x \rangle^2_{\tilde{q},a,\beta}} & = &   
\frac{ \mbox{B}(3 \beta,1/\tilde{q} - 3 \beta) \mbox{B}( \beta,1/\tilde{q} -  \beta)}
     {\mbox{B}^2(2 \beta,1/\tilde{q} - 2 \beta)} - 1 \; . 
\end{eqnarray}
Notice that the ration $\sigma_{\tilde{q},a,\beta}^2/\langle x \rangle^2_{\tilde{q},a,\beta}$ depends only on $\beta$ and $\tilde{q}$, but not on $a$. 

Particular values of $\beta$ leads to a $\tilde{q}$-generalization of the error function and to the Zipf-Mandelbrot pdf. 

\subsubsection{Generalized Error Function.}

To generalize the error function, consider $\beta = 1/2$ and $a = \beta^{\beta}$ (or $\alpha = 1$) in Eq.~\ref{eq:cumulative_generalized_strechted_exponential} and one has:
\begin{eqnarray}
\mbox{erf}_{\tilde{q}}(x) & = & F_{\tilde{q},1/\sqrt{2},1/2}^{(1)}(x) 
                                       =  \frac{ 2 \sqrt{\tilde{q}} }{ \mbox{B}(1/2,1/\tilde{q} - 1/2)} \; \int_0^{x} dt \;  e_{-\tilde{q}}(-t^2)\; .
\label{eq:generalized_error_function_1}
\end{eqnarray}
As $\tilde{q} \to 0$, $\mbox{B}(1/2,1/\tilde{q} - 1/2) = \sqrt{\tilde{q} \pi}$ and one retrieves the standard error function: $\mbox{erf}_{0}(x) = \mbox{erf}(x) = (2/\sqrt{\pi}) \int_0^x dt \; e^{-t^2} $.

\subsubsection{Zipf-Mandelbrot PDF.}

For $\beta = 1$ in Eq.~\ref{eq:generalized_strechted_exponential}, one obtains the Zipf-Mandelbrot's pdf:
\begin{equation}
P_{\tilde{q}, \alpha, 1}^{(1)}(x) = \frac{1 - \tilde{q}}{\alpha} \; e_{-\tilde{q}}\left(\frac{-x}{\alpha} \right) \frac{1 - \tilde{q}}{\alpha} \; \frac{1}{(1 + \tilde{q} x/\alpha)^{1/\tilde{q}}} \; .
\label{eq:zipf_mandelbrot_pdf}
\end{equation}
which mean value and variance are:
\begin{eqnarray}
\langle x \rangle_{\tilde{q}, \alpha, 1} & = & \frac{\alpha}{1 - 2 \tilde{q}} \\ 
     \frac{\sigma_{\tilde{q}, \alpha, 1}^2}
          {\langle x \rangle^2_{\tilde{q}, \alpha, 1}} & = & \frac{\tilde{q}-1}{3\tilde{q}-1} \; , 
\end{eqnarray}
which is finite for $0 \le \tilde{q} < 1/3$ and, from Eq.~\ref{eq:cumulative_generalized_strechted_exponential}, one obtains its cumulative function: 
\begin{eqnarray}
F_{\tilde{q},a,1}^{(1)}(x) & = & 1 - \frac{1}{[e_{\tilde{q}}(x/a)]^{1-\tilde{q}}} =1 - \tilde{F}_{\tilde{q},a,1}^{(1)}(x) \; , 
\end{eqnarray}
where the upper-tail distribution its simply given by:
\begin{eqnarray}
\nonumber
\tilde{F}_{\tilde{q},a,1}(x) & = & \int_x^{\infty} dt \; P_{\tilde{q},a,1}^{(1)}(t) = \left(1 + \tilde{q} \; \frac{x}{a} \right) \; e_{-\tilde{q}} \left( \frac{- x}{a} \right) \\
             & = & \left[ e_{-\tilde{q}} \left( \frac{- x}{a} \right) \right]^{1-\tilde{q}} =  \left[ e_{\tilde{q}} \left( \frac{ x}{a} \right) \right]^{\tilde{q}-1}\; ,
\label{eq:uppertail_zipf_mandelbrot}
\end{eqnarray}
which is a more suitable for fitting the model to real data than the pdf (Eq.~\ref{eq:zipf_mandelbrot_pdf}) itself. 

\subsection{Two Tail PDF}

If the domain of the considered independent variable is not bounded, it is interesting to consider its  absolute value $|x|$ in  Eq.~\ref{eq:generalized_strechted_exponential} and one has a symmetric pdf about the line $x=0$. 
Notice that in this case, the normalization factor must be halved since the domain has been doubled in a symmetrical way, the pdf is then: 
\begin{equation}
P_{\tilde{q},a,\beta}^{(2)}(x) = \frac{ ( \tilde{q} \beta)^{\beta}}{2 a \beta \; \mbox{B}(\beta,1/\tilde{q} - \beta)}  \;  e_{-\tilde{q}} \left[ - \beta \; \left(\frac{|x|}{a}  \right)^{1/\beta} \right] \; ,
\label{eq:generalized_strechted_exponential_pdf}
\end{equation}
which is a convenient function to use in wavelets~\cite{borges_2004:b}. 
Its cumulative function is:
\begin{equation}
F_{\tilde{q},a,\beta}^{(2)}(x) = \frac{1}{2} \left[ 1+ \mbox{sgn}(x) F_{\tilde{q},a,\beta}^{(1)}(|x|) \right] \; ,
\end{equation}
where $F_{\tilde{q},a,\beta}^{(1)}(x)$ is given by Eq.~\ref{eq:cumulative_generalized_strechted_exponential}.

On one hand, due to its symmetry around $x = 0$, the odd moments of this pdf vanish $\langle x^{2n + 1} \rangle = 0$, with $n = 0,1,2, \ldots$. 
On the other hand, the even moments $\langle x^{2n} \rangle$ are finite only if $\tilde{q} < 1/[(2n + 1)\beta]$: 
\begin{equation}
\langle x^{2n} \rangle_{\tilde{q},a,\beta} = \left[ \frac{a}{(\tilde{q} \beta)^{\beta}} \right]^{2n}
\frac{  \mbox{B}[(1 + 2n) \beta,1/\tilde{q} - (1 + 2n) \beta]}
     { \mbox{B}( \beta,1/\tilde{q} -  \beta)}    \; .
\label{eq_even_moments}
\end{equation}

In the following we retrieve the generalized gaussian as a particular case of  Eq.~\ref{eq:generalized_strechted_exponential_pdf}.
Also, we propose to consider another particular case of Eq.~\ref{eq:generalized_strechted_exponential_pdf} to generalize the laplacian pdf. 
The characteristic function of both particular cases are analytically calculated. 
  
\subsubsection{Generalized Gaussian.}

An interesting particular case of the two-tail generalized stretched exponential function is when $\beta = 1/2$, which leads to the $\tilde{q}$-guassian~\cite{PhysRevLett.75.3589}: 
\begin{eqnarray}
G_{\tilde{q},a}(x) & = &  \frac{\sqrt{\tilde{q}} \; \Gamma(1/\tilde{q})}{\Gamma(1/\tilde{q} - 1/2)} \; \frac{e_{-\tilde{q}}[-(x/a)^{2}/2]}{\sqrt{2 \pi a^2}}  \; .
\label{eq:qgaussian}
\end{eqnarray}
Due to symmetry, all odd moments vanish.
Even moments are given by Eq.~\ref{eq_even_moments}:  
\begin{equation}
\langle x^{2n} \rangle_{\tilde{q},a,1/2} = \left( \frac{2 a^2}{\tilde{q}} \right)^{n}
\frac{ \Gamma(1/2 + n) \Gamma(1/\tilde{q} - 1/2 - n) }
     { \sqrt{\pi} \; \Gamma( 1/\tilde{q} -  1/2)}    \; .
\end{equation}
and the variance is finite only for $\tilde{q} < 2/3$: 
\begin{equation}
\sigma_{\tilde{q},a}^2 = \langle x^2 \rangle_{\tilde{q},a} = \frac{2 a^2}{2 - 3 \tilde{q}}\; .
\end{equation}

Using that for $a \gg b$, $\Gamma(a + b)/\Gamma(a) = a^b$,  when $\tilde{q} \rightarrow 0$ in Eq.~\ref{eq:qgaussian}, $\Gamma(1/\tilde{q})/\Gamma(1/\tilde{q} - 1/2) = 1/\tilde{q}^{1/2}$ and one has a guassian, with variance $a^2$: $G_{0,a}(x) = e^{-(x/a)^2/2}/\sqrt{2 \pi a^2}$,
as a particular case. 
Another particular case is when $\tilde{q} = 1$ and one retrieves the Lorentzian (Cauchy pdf)
$G_{1,a}(x) = 1/\{\pi a \sqrt{2 }[1 + (x/a)^2/2]\}$.

The characteristic function of Eq.~\ref{eq:qgaussian} has an analytical closed form:
\begin{eqnarray}
\nonumber
p_{\tilde{q},a, 1/2}(k) & = & \langle e^{\imath k x} \rangle = \int_{-\infty}^{\infty} dx \; G_{\tilde{q},a}(x) e^{\imath k x} \\
                   & = & \left( \frac{ ka\sqrt{2/\tilde{q}}}{2} \right)^{1/\tilde{q} - 1/2} \frac{2 K_{1/\tilde{q} - 1/2}(ka\sqrt{2/\tilde{q}})}{\Gamma(1/\tilde{q} - 1/2)} \; ,
\end{eqnarray}
where $K_{\nu}(z)$ is the $K$-modified Bessel function~\cite{stegun}: 
\begin{equation}
K_{\nu}(z)  =  \frac{2^{\nu} \Gamma(\nu + 1/2)}{\sqrt{\pi} z^{\nu + 1}} \; \int_0^{\infty} \frac{dt \cos t}{[1 + (t/z)^2]^{\nu + 1/2}} \; . 
\end{equation}
For the Gaussian one has also a Gaussian $p_{0,a}(k) = e^{- a^2 k^2/2}$ but for the Lorentzian, one has the Laplace function $p_{1,a} = e^{\sqrt{2}a |k|}$. 

\subsubsection{Generalized Laplace PDF.}

For $\beta = 1$ and $\tilde{q} = 0$, Eq.~\ref{eq:generalized_strechted_exponential_pdf} leads  to the Laplace pdf $P_{0,a,1}^{(2)}(x) = e^{-|x|/a}/(2 a)$. 
For arbitrary $\tilde{q}$, one has the \emph{generalized Laplace pdf}: 
\begin{equation}
L_{\tilde{q},a}(x)  = P_{\tilde{q},a,1}^{(2)}(x) = \frac{1 - \tilde{q}}{2a} \; e_{-\tilde{q}}\left( \frac{-|x|}{a} \right) \; , 
\label{eq:qlaplace}
\end{equation}
and its cumulative function is:
\begin{equation}
F_{\tilde{q},a,1}^{(2)}(x) = \frac{1}{2} \tilde{P}_{\tilde{q},a,1}(|x|)
\; ,
\end{equation}
where $\tilde{P}_{\tilde{q},a,1}(x)$ is given by Eq.~\ref{eq:zipf_mandelbrot_pdf}.

The odd moments of Eq.~\ref{eq:qlaplace} vanish and the even ones are finite if $\tilde{q} < 1/(1 + 2n)$:
\begin{equation}
\langle x^{2n} \rangle_{\tilde{q},a,1} = \frac{1}{\tilde{q} - 1} \; \left( \frac{a}{\tilde{q}} \right)^{2n} \; \mbox{B}(1 + 2n, 1/\tilde{q}-2n -1)   \; .
\end{equation}
Its characteristic function has an analytical closed form:
\begin{eqnarray}
\nonumber
p_{\tilde{q},a,1}(k) & = & \int_{-\infty}^{\infty} dx \; L_{\tilde{q},a}(x) e^{\imath k x} 
= \frac{1 - \tilde{q}}{a} \; \int_0^{\infty} dx \; \cos(kx) \; e_{-\tilde{q}}(-x/a) \\
\nonumber
                   & = & \frac{1 - \tilde{q}}{a} \; \int_0^{\infty} dx \; \frac{\cos(kx)}{(1 + \tilde{q} x/a)^{1/\tilde{q}}} \\
                   & = &  \frac{\pi \, t_{0}^{1 / \tilde{q} - 1} \, \sin(t_{0} + \pi / (2 \tilde{q}))}
{\sin(\pi / \tilde{q}) \Gamma(1 / \tilde{q} - 1)} +  _{1}F_{2} \left(1; 1 - \frac{1}{2 \tilde{q}}, \frac{3}{2} - \frac{1}{2 \tilde{q}}; \frac{-t_{0}^{2}}{4} \right) \; .
                    \end{eqnarray}
where $t_0 = k a/\tilde{q}$ and $_{1}F_{2}(a;b_1,b_2;z) = \sum_{n=0}^{\infty} \{[(a)_n]/[(b_1)_n (b_2)_n]\} (x^n)/n!$ is the hypergeometric function~\cite{stegun}, with $(a)_n$ being the Pochhammer symbol.

\section{Conclusion}
\label{sec:conclusion}

We have shown that the $\tilde{q}$-generalization of the exponential is suitable to describe complex systems since it generalizes functions, which govern these systems.  
As special cases we have obtained the $\tilde{q}$-generalized stretched exponential function, which has the generalized error function,  
 the generalized Laplace pdf and the already known generalized gaussians as special cases.  
Further, we have used the $\tilde{q}$-exponential to write the very reliable rank distribution obtained by Naumis and Cocho. 
An interesting application of the $\tilde{q}$-exponential is in the generalization of the gamma function, which allows us to obtain a new generalization for the factorial operation. 
Since these distributions are the solution of differential equations that describe the complex systems, the $\tilde{q}$ generalization brings many different systems to be described by the same underlying process.

\section*{Acknowledgments}



ASM acknowledges the Brazilian agencies CNPq (303990/2007-4 and 476862/2007-8) for support. 
RSG also acknowledge CNPq (140420/2007-0) for support. 


\end{document}